\documentclass[12pt,aps,showpacs]{revtex4}
\usepackage{graphicx}
\begin{document}
\title{Creation of entanglement in a scalable spin quantum computer \\ with
long-range dipole-dipole interaction between qubits}
\author{D. I. Kamenev$^{1}$, G. P. Berman$^1$,
and V. I. Tsifrinovich$^{2}$}
\affiliation{$^1$Theoretical Division, T-13, and the Center for
Nonlinear Studies, Los Alamos National Laboratory, Los Alamos, New
Mexico 87545}
\affiliation{$^3$ IDS Department, Polytechnic University, Brooklyn,
New York 11201}

\begin{abstract}
Creation of entanglement is considered theoretically and numerically
in an ensemble of spin chains with dipole-dipole interaction between
the spins in each chain. The unwanted effect of the long-range
dipole interaction is compensated by the optimal choice
of the parameters of radio-frequency pulses implementing the
protocol. The errors caused by (i) the influence of the environment,
(ii) non-selective excitations, (iii) influence of different spin
chains on each other, (iv) displacements of qubits from their
perfect locations, and (v) fluctuations of the external magnetic
field are estimated analytically and calculated numerically. For the
perfectly entangled state the $z$ component, $M$, of the
magnetization of the whole system is equal to zero. The errors lead
to a finite value of $M$. If the number of qubits in the system is
large, $M$ can be detected experimentally. Using the fact that $M$
depends differently on the parameters of the system for each kind of
error, varying these parameters would allow one to experimentally
determine the most significant source of errors and to optimize
correspondingly the quantum computer design in order to decrease the
errors and $M$. Using our approach one can benchmark the quantum
computer, decrease the errors, and prepare the quantum computer for
implementation of more complex quantum algorithms.
\end{abstract}



\maketitle
\section{Introduction}
A quantum computer is supposed to be an analog device designed for
implementation of quantum  algorithms. The most important algorithm
is the Shor's quantum algorithm for factorization of large integer
numbers~\cite{Shor,book}. Before experimental realizatation of this
and other complex quantum algorithms, this computer must be tested
by implementation of simple algorithms, such as, for example,
creation of entanglement with two quantum states (and many qubits).
This procedure would allow one to identify the most significant
sources of errors and to optimize the quantum computer design.

The following criteria~\cite{criteria} must be met by a physical
implementation of a quantum information processor: (i) the qubits
must be easy to physically manipulate, (ii) easy to increase the
number of qubits, (iii) qubits should interact with each other, (iv)
the qubits should be somewhat isolated from environment, and (v) the
qubit states must be detectable. Currently, solid state
implementations of a quantum computer allow (theoretical)
scalability [criterion (ii)] and possibility to substantially
increase the number of qubits in the quantum computer register. In a
solid state currently one and two-qubit quantum logic operations are
implemented and measured experimentally. Nuclear magnetic resonance
and electron spin resonance techniques have been used for quantum
information processing in solids. Etanglement between electron and
nuclear spins of the same molecule has been implemented and measured
in a malonic acid single crystal~\cite{refMolecular} and in
$^{15}$N@C$_{60}$ endohedral fullerene~\cite{refFullerene}. Quantum
process tomography has been performed on a solid-state qubit
represented by a nitrogen-vacancy defect in
diamond~\cite{refTomography} at room temperature.  Initialization
(cooling) of nuclear spin has been realized in isotopically labeled
malonic acid molecules by a controllable transfer of polarization
from neighboring nuclear spins~\cite{refCooling,refMalonicAcid} at
room temperature. One and two-qubit quantum logic operations have
been implemented and measured in superconducting quantum
computers~\cite{Ex5,Ex10,1shotMeasurement}. A further technological
advance can be implementation of simple quantum computing algorithms
in a scalable solid state system with a large (more than two) number
of qubits for demonstration of basic principles of quantum
computation. In this paper, we consider how to implement, probably,
the simplest possible algorithm with many qubits: creation of
entanglement. To implement this algorithm, we choose one of the most
affordable systems: a spin chain placed in a permanent magnetic
field with a gradient along the chain. There is a constant
dipole-dipole interaction between the qubits. The logic operations
are implemented by rectangular radio-frequency pulses. This setup
allows one to achieve all criteria (i)-(v).

Indeed, criterion (i) is satisfied because in the system there are
no switchable interactions controlled by nanoscopic metal gates, and
quantum logic operations are implemented using global addressing
technique based on rectangular radio-frequency pulses. Criterion
(ii) is satisfied because we consider solid-state QC architectures
and because rectangular pulses can be used in the QC with many
qubits and all parameters of the applied pulses can be determined
analytically. The number of pulses in our entanglement protocol is
equal to the number of qubits in the chain. The long-range constant
magnetic dipole-dipole interaction satisfies criterion (iii). In the
majority of spin-based quantum computer architectures this is the
only interaction when neighboring qubits are separated from each
other at the distances $>1$ nm. Criterion (iv) is satisfied because
our system allows implementations using qubits with long decoherence
times. One possible implementation is based on nuclear spins, for
example nuclear spins of $^{31}$P~\cite{Kane},
$^{29}$Si~\cite{all-silicon}, or Li~\cite{Li} in $^{28}$Si. Another
implementation is based on endohedral fullerenes, $^{15}$N@C$_{60}$
and $^{31}$P@C$_{60}$~\cite{fulleren02}, where the fulleren cage
provides a good isolation of electron or nuclear spin of the
enclosed atom of nitrogen or phosphorus from the environment. The
electrons of the nitrogen or phosphorus atoms have spin 3/2, and the
nuclei have spin 1/2~\cite{fullerenEx1,fullerenEx2}. Consequently,
our protocols can work only for the nuclear spins of the
endohedrals. However, a modification of our scheme for the electrons
with spin 3/2 is possible~\cite{fulleren03}. The third possible
implementation is based on electron spins in self-assembled
monolayer systems~\cite{Allara,Allara1,Allara2}. Measurement
[criterion (v)] can be performed by using an ensemble of many
identical spin chains~\cite{Allara} shown in Fig.~1. The external
magnetic field in Fig.~1 is nonuniform in the $x$ direction but
uniform in the $y$ direction. Such a gradient can be created by
current(s) flowing along the $y$ axis in the plain formed by the
chains. Consequently, we take the angle $\Theta$ between the qubit
plane and the permanent magnetic field to be $\Theta=\pi/2$. If the
distance between neighboring chains is sufficiently large all chains
experience the same conditions. Consequently, one can create
entanglement simultaneously in all chains. If the number of spins in
the system is sufficiently large, one can detect the macroscopic
magnetization $\vec{\cal M}$ of the whole system. If each chain is
in the entangled state
\begin{equation}
\label{entangled}
{1\over \sqrt 2}\left(|00\dots 00\rangle+
e^{i\varphi}|11\dots 11\rangle \right),
\end{equation}
then $z$ component ${\cal M}_z$ of the macroscopic magnetization
must be equal to zero. All possible errors result in deviation of
${\cal M}_z$ from its perfect value ${\cal M}_z=0$. We will show in
this paper that the sign of ${\cal M}_z$ and the dependence of
${\cal M}_z$ on different parameters of the model can provide us the
information about the main source of error. This information can be
useful for benchmarking the quantum computer and optimization of its
architecture and parameters of quantum protocols.

\begin{figure}[t!]
\vspace{-30mm}
\centerline{\includegraphics[width=14cm,height=14cm,angle=270,clip=]{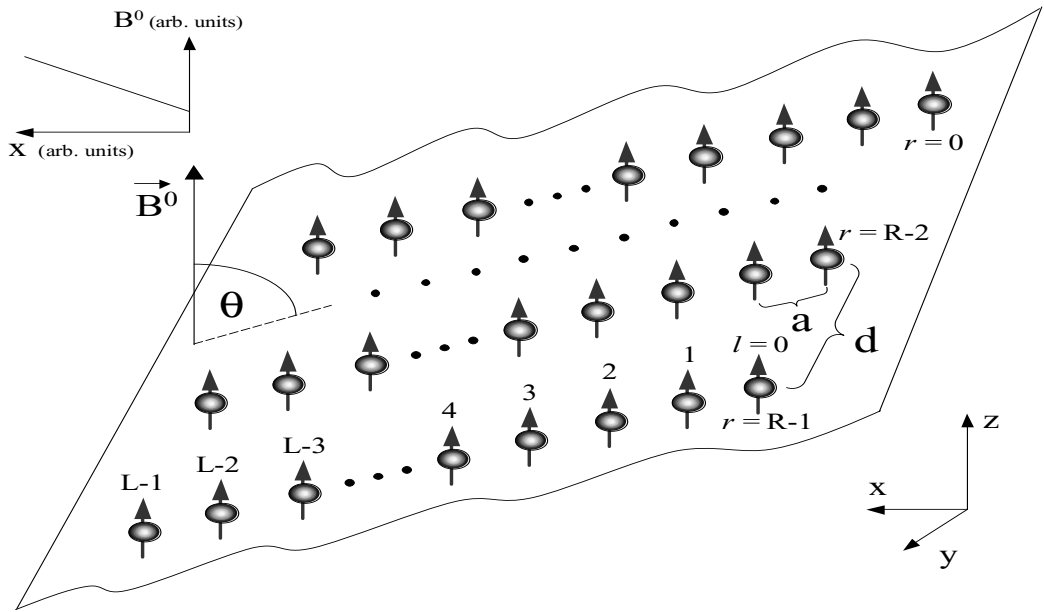}}
\vspace{-27mm} \caption{A setup for creation of entanglement with
ensemble of spin chains.}
\label{reffig:fig1}
\end{figure}

\section{Hamiltonian}
We consider the implementation of entanglement
only for one spin chain of the ensemble of chains illustrated in
Fig.~1. Since the chains experience the same conditions
the dynamics of all chains is similar. One can neglect
the magnetic dipole-dipole interaction between the spins of the
different chains when the condition $(a/d)^3\ll 1$ is satisfied.
(Here $a$ is the distance between neighboring qubits of
a single chain and $d$ is the distance between different chains,
see Fig. 1.) More exact condition of relative independence of
different chains will be given in Section~\ref{sec:differentChains}
below.

The applied magnetic field has the following components
$$
\vec B_n(x,t)=\left[B^0(x),B^1_n\cos(\nu_n t+\varphi_n),
-B^1_n \sin(\nu_n t+\varphi_n)\right],
$$
where $B^0(x)$ is a permanent magnetic field and
$B^1_n$, $\nu_n$, and $\varphi_n$ are, respectively,
the amplitude, frequency, and phase of the $n$th
circularly polarized radio-frequency rectangular pulse of a protocol.
The permanent magnetic field $B^0(x)$ has a constant gradient
$B^0(x_l)=\omega_0+l\delta B^0$, where $x_l$ is the $x$-coordinate
of the $l$th qubit. Note that in practice the magnetic field
gradient can be variable. Our results can be easily
reformulated for this case, provided that
$|B^0(x_l)-B^0(x_{l-1})|\gg B^1_n$.
If there is the dipole-dipole interaction
between the qubits the Hamiltonian is
\begin{equation}
\label{Hn}
H_n=H^0+H_{\rm int}+V_n(t),
\end{equation}
where
\begin{equation}
\label{H0V}
H^0=-\sum_{l=0}^{L-1}\omega_lS_l^z,\qquad
V_n(t)=-{\Omega_n\over 2}\sum_{l=0}^{L-1}
\left\{S_l^-\exp\left[-i\left(\nu_n t+
\varphi_n\right)\right]+h.c.\right\},
\end{equation}
\begin{equation}
\label{Hint}
H_{\rm int}=-{J\over A^3}
\sum_{l=0}^{L-1}\sum_{k=l+1}^{L-1}{1\over (k-l)^3}
S_l^zS_k^z,
\qquad
J={\mu^2\over \hbar {\cal A}^3}(3\cos^2\Theta-1)=
-{\mu^2\over \hbar {\cal A}^3}.
\end{equation}
Here the Hamiltonian is presented in the frequency units; $\hbar$ is
the Planck constant; $S_l^\pm=S_l^x\pm iS_l^y$; $S_l^x$, $S_l^y$,
and $S_l^z$ are the components of the operator of the $k$th nuclear
or electron spin $1/2$; $\omega_l=\omega_0+l\delta\omega$,
$\omega_0=\gamma B^0(x_0)$, $\delta\omega=\gamma\delta B^0$;
$\Omega_n=\gamma B^1_n$ is the Rabi frequency of the $n$th pulse;
$\gamma=\gamma_N$ or $\gamma=-\gamma_e$, where $\gamma_N$ and
$\gamma_e$ are, respectively, the nuclear and electronic
gyromagnetic ratios; ${\cal A}=1$ nm, $A$ is the dimensionless
parameter equal to the distance between neighboring qubits measured
in nanometers, so that $a=A{\cal A}$, ($a$ is the distance between
neighboring qubits); $\Theta=\pi/2$ is the angle between direction
of the spin chain and direction of the permanent magnetic field;
$\mu=-g_e\mu_{\rm B}$ for an electron spin and $\mu=g_N\mu_N$ for a
nuclear spin, $\mu_{\rm B}$ and $\mu_N$ are, respectively, the Bohr
and nuclear magnetons, $g_e\approx 2$ and $g_N$ are, respectively,
the electron and nuclear $g$-factors. In Eq.~(\ref{Hint}) we neglect
the $x$ and $y$ components of the dipole-dipole interaction because
the magnetic dipole field on $l$th qubit in any stationary state
$|00\dots 00\rangle$, $|00\dots 01\rangle$, \dots is much less than
$\delta B^0$. So, only the $z$ component of the dipole-dipole field
significantly affects the quantum dynamics.

\section{Creation of entanglement}
\label{sec:entanglement} First, we discuss the formal steps required
to create the entangled state. Let the initial state be the ground
state $|0_{L-1}0_{L-2}\dots 0_10_0\rangle$ as shown in Fig.~1. We
split the ground state into two states by applying the Hadamard
transformation
\begin{equation}
\label{VH} {\cal H}_0|0_{L-1}0_{L-2}\dots 0_10_0\rangle=
{1\over\sqrt 2}(|0_{L-1}0_{L-2}\dots 0_10_0\rangle
+e^{i\phi^\prime}|0_{L-1}0_{L-2}\dots 0_11_0\rangle).
\end{equation}
Here and below we omit the total phase factor.
Then we apply Control-Not gate CN$_{01}$ to the 1st qubit
to obtain
\begin{equation}
\label{2ndStep}
{1\over\sqrt 2}(|0_{L-1}0_{L-2}\dots 0_10_0\rangle
+e^{i\phi^{\prime\prime}}|0_{L-1}0_{L-2}\dots 0_21_11_0\rangle).
\end{equation}
Applying Control-Not gates to the remaining $L-2$ qubits we
obtain the entangled state:
\begin{equation}
\label{finalState}
{1\over\sqrt 2}(|0_{L-1}0_{L-2}\dots 0_10_0\rangle
+e^{i\phi}|1_{L-1}1_{L-2}\dots 1_11_0\rangle).
\end{equation}
Below we do not take into consideration the phase factor
$e^{i\phi}$. One can make $\phi$ to be equal to zero by
a proper choice of the phases $\varphi_n$ of the
pulses~\cite{Method}.

\section{Quantum dynamics}
We decompose the wave function into the basis states $|p\rangle$
of the unperturbed Hamiltonian $H^0$:
\begin{equation}
\label{psi}
\psi(t)=\sum_{p=0}^{2^L-1}C_p(t)e^{-iE_pt}|p\rangle,~~~
\end{equation}
where $|p\rangle=|n_{L-1}n_{L-2}\dots n_l\dots n_1n_0\rangle$,
$n_l=0,1$, and
\begin{equation}
\label{energy}
E_p=\langle p|H^0+H_{\rm int}|p\rangle=
-\sum_{l=0}^{L-1}\omega_l(p)s^z_l(p)-{J\over A^3}\sum_{l=0}^{L-1}
\sum_{k=l+1}^{L-1}{s^z_l(p)s^z_k(p)\over (k-l)^3}.
\end{equation}
Here $s^z_l(p)$ is the eigenvalue of the operator $S^z_l$ in the
state $|p\rangle$: if the $l$th spin of the state $|p\rangle$ is in
the state $|\dots 0_l\dots\rangle$ then $s^z_l(p)=1/2$ and if the
$l$th spin of the state $|p\rangle$ is in the state $|\dots
1_l\dots\rangle$ then $s^z_l(p)=-1/2$.

Let the $j$th spin of the state $|p\rangle$ be in the state
$|\dots 0_j\dots\rangle$ and let the state $|q\rangle$ be
associated with the state $|p\rangle$ by flip of the
$j$th spin, $|q\rangle=|\dots 1_j\dots\rangle$. If
the frequency $\nu_n$ of the electromagnetic field
is close to the Larmor frequency of the $j$th spin, then
under the conditions $|\delta\omega|\gg J/A^3$ and
$|\delta\omega|\gg\Omega_n$
the pulse affects mostly this $j$th spin, and the
approximate solution is~\cite{Lopez00}
$$
C_p(t_n+\tau_n)=\left\{\cos\left[{\lambda_n(q,p)\tau_n\over 2}\right]+
i{\Delta_n(q,p)\over\lambda_n(q,p)}
\sin\left[{\lambda_n(q,p)\tau_n\over 2}\right]\right\}
e^{-i\Delta_n(q,p)\tau_n/2},
$$
\begin{equation}
\label{2x2}
C_q(t_n+\tau_n)=i{\Omega_n\over\lambda_n(q,p)}
\sin\left[{\lambda_n(q,p)\tau_n\over 2}\right]
e^{i\Delta_n(q,p) t_n-i\varphi_n+i\Delta_n(q,p)\tau_n/2},
\end{equation}
$$
C_p(t_n)=1,\qquad C_q(t_n)=0,
$$
where $t_n$ is the time of the beginning of the $n$th pulse and
$$
\Delta_n(q,p)=E_q-E_p-\nu_n,\qquad
\lambda_n(q,p)=\sqrt{\Delta^2_n(q,p)+\Omega_n^2}.
$$
If the detuning $\Delta_n(q,p)$ is equal to zero, $\Delta_n(q,p)=0$,
and for $\Omega_n\tau_n=\pi$ ($\pi$-pulse), there is a complete transition
between the states $|p\rangle$ and $|q\rangle$.
Here we neglected the transitions associated with flips
of spins with nonresonant transition frequencies for which
$l\ne j$. The corrections to the probability amplitude
associated with flips of these spins are of order of
$\Omega_n/(2|\delta\omega||j-l|)$~\cite{PRA01,JAM}. These corrections
are small provided $\Omega_n/|\delta\omega|\ll 1$.

\subsection{Control-Not gate for the system with long-range
dipole-dipole interaction} Since the entanglement is implemented by
a sequence of Control-Not gates we now derive the parameters
required to implement these gates taking into consideration the
long-range interaction. Because of this interaction the action of
the Control-Not gate on the $j$th qubit depends not only on the
states of $(j-1)$th and $(j+1)$th qubits, but also on the states of
distant $l$th qubits, where $l\ne j-1,j,j+1$. Consequently, it is
convenient to formulate the Control-Not gate CN not in terms of the
states of the corresponding qubits, like CN$_{j-1,j}$, but in terms
of the eigenstates $|p\rangle$ of the Hamiltonian $H^0$. For
example, the gate CN$_j(p,q)$ flips the $j$th qubit for the state
$|p\rangle$ and completely suppresses flip of the same $j$th qubit
for the state $|q\rangle$. This procedure allows one to compensate
the unwanted action of the long-range interaction by an optimal
choice of the parameters of the pulses. Our approach works for the
case when there are only two ``working states'' (excluding error
unwanted states) in the quantum register. It is useful for creation
of entanglement with two states introduced in
Section~\ref{sec:entanglement}. If there are more than two states in
the quantum register, more complex (shaped) pulses are required to
minimize the effect of the long-range interaction.

Let us write the entangled state in the form
$(1/\sqrt 2)(|0\rangle+|p^\prime\rangle)$, where $|0\rangle$ is the
ground state and $|p^\prime\rangle$ is the excited state. The
pulses of the protocol must suppress the transitions from the
ground state and to implement the transitions for the excited states,
so that the state $|p^\prime\rangle$ evolves as
$$
|p^\prime\rangle:~|000\dots001\rangle\rightarrow |000\dots011\rangle
\rightarrow |000\dots0111\rangle\rightarrow \dots
\rightarrow
$$
\begin{equation}
\label{excited}
|001\dots111\rangle\rightarrow
|011\dots 11\rangle\rightarrow |111\dots 11\rangle.
\end{equation}
The Control-Not gates implementing these transitions
are CN$_j(p^\prime,0)$, where $j=1,2,\dots,L-1$ is the number
of the spin to be flipped in the excited state $|p^\prime\rangle$
and also the number of the corresponding $\pi$-pulse.
(We do not count the initial $\pi/2$ pulse.)

We now derive the parameters of the initial pulse
implementing the Hadamard transform.
In order to flip the $0$th spin in the ground state
the detuning must be equal to zero, $E_1-E_0-\nu_{H}=0$,
where $\nu_{H}$ is the frequency of the pulse.
Using Eq.~(\ref{energy}) we obtain
\begin{equation}
\label{frequencyH}
\nu_{H}=E_1-E_0=
\omega_0+{J\over 2 A^3}\sum_{l=1}^{L-1}{1\over l^3}.
\end{equation}
The Rabi frequency $\Omega_{H}$ of the initial pulse must satisfy
the condition $\Omega_{H}\ll\delta\omega$
and the time-duration of the pulse is
$\tau_{H}=\pi/(2\Omega_{H})$.

Consider now the Control-Not gate CN$_j(p^\prime,0)$.
If the state $|q^\prime\rangle$ is associated with the state
$|p^\prime\rangle$ by a flip of the $j$th spin, the
frequency is
\begin{equation}
\label{frequency}
\nu_j=E_{q^\prime}-E_{p^\prime}=\omega_j+
{J\over A^3}\sum_{l=0\atop l\ne j}^{L-1}{s^z_l(p^\prime)\over |l-j|^3}.
\end{equation}
For example, for $j=1$, one has
$s^z_0(p^\prime)=-1/2$ and
$s^z_l(p^\prime)=1/2$ for $l=2,3,\dots,L-1$,
so that the frequency of the first $\pi$ pulse is
$$
\nu_1=\omega_1+{J\over 2A^3}\sum_{l=3}^{L-1}{1\over (l-1)^3};
$$
the frequency of the second $\pi$ pulse is
$$
\nu_2=\omega_2+{J\over 2A^3}\sum_{l=5}^{L-1}{1\over (l-2)^3},
$$
and so on.

In order to suppress the unwanted transitions from the
ground states one can use a $2\pi K$ method~\cite{book,book1,1997}.
Here we will modify this method so that it can be applied
to the system with the long-range interaction.
As follows from Eq.~(\ref{2x2}) the
transition with nonzero detuning is suppressed
if the value of the sine is equal to zero, i.e. when
\begin{equation}
\label{2piK}
\Omega_j={|\Delta_j(q^{\prime\prime},0)|\over \sqrt{4K^2-1}},
\end{equation}
where $K=1,2,\dots$ and the state $|q^{\prime\prime}\rangle$ is associated
with the ground state by a flip of the $j$th spin.
Using Eq.~(\ref{energy}) we find
$$
E_{q^{\prime\prime}}-E_0=\omega_j+{J\over A^3}
\sum_{l=0\atop l\ne j}^{L-1}{s_l^z(0)\over |l-j|^3}=\omega_j+
{J\over 2A^3}\sum_{l=0\atop l\ne j}^{L-1}{1\over |l-j|^3}.
$$
From Eq.~(\ref{frequency}) we obtain the detuning
\begin{equation}
\label{detuning}
\Delta_j(q^{\prime\prime},0)=E_{q^{\prime\prime}}-E_0-\nu_j={J\over A^3}
\sum_{l=0\atop l\ne j}^{L-1}{s_l^z(0)-s_l^z(p^\prime)\over |l-j|^3}=
{J\over A^3}\sum_{l=0}^{j-1}{1 \over (j-l)^3}.
\end{equation}
We note that if the long-range dipole-dipole interaction
is not taken into consideration, then the error to the
probability amplitude generated
by each pulse due to the next-nearest neighbor interaction is
of the order of $(1/2)^3$=$1/8$. Our approach allows
to compensate the unwanted effect of the long-range interaction
by optimal choice of the parameters of the pulses.

\section{Errors and $z$ component of the magnetization}
The protocol consisting of rectangular pulses with the
parameters defined by Eqs.~(\ref{frequencyH})-(\ref{detuning})
allows one to implement entanglement in the ensemble of
noninteracting spin chains with minimum possible error $P$.
It is possible to relate the error $P$ with the
$z$ component ${\cal M}_z$ of the magnetization of the system.
Since ${\cal M}_z$ is caused by the error states, then one can assume
that ${\cal M}_z\sim P$. We define the dimensionless $z$ component of
the magnetization $M\sim {\cal M}_z$ in the following way. Let the
maximum value of $|M|$ be 1. Then for the state $|00\dots 00\rangle$
we have $M=1$ and for the state $|11\dots 11\rangle$
we have $M=-1$. The measured z-component of the magnetization
of the system shown in Fig.~1 corresponding to $M$ is
${\cal M}_z={1\over 2}M\mu RL$.

In this Section we will estimate and compare contribution
of different kinds of errors to the total error $P$ and relate these
errors with $M$. Since we can estimate $P$ analytically for a
large number of qubits, the relation between $M$ and $P$
allows us to estimate $M$ for $L\gg 1$.
Different kinds of errors can generate
positive and negative contributions to $M$ so that they can
cancel each other. For example, the state
$C_p(|000111\rangle+|111000\rangle)$ contributes to $P$
but does not contribute to $M$. In spite of this we
assume that for a definite range of parameters
the relation $M=gP$ holds [see Eq.~(\ref{MP}) below],
where the coefficient $g$ is the ``geometrical factor'' and $P$ is the probability
of error. For example, if a single spin in the state $(1/\sqrt 2)|0000\rangle$
of superposition (\ref{entangled})
is flipped down with the probability $P$ then $g=-1/L=-1/4$.
If two spins are flipped down with the probability $P$ then $g=-2/L=-1/2$.
If a single spin in the state $(1/\sqrt 2)|1111\rangle$
is flipped up then $g=1/L=1/4$ and so on. If, for example, different spins
in the state $(1/\sqrt 2)|0000\rangle$ have different probabilities
$P_j$ to be flipped then $M=(-1/L)\sum_{j=0}^3P_j$
[see Eqs.~(\ref{Md}) and (\ref{Mosc}) below].

The probability of error
$P$ depends on parameters and number of qubits $L$. This
dependence is different for different kinds of errors. For example,
for nonresonant transitions $P\sim L$ [see Eq.~(\ref{P_nr}) below].
By experimental measurement of $M$
for different parameters one can define the most important
mechanism responsible for the errors. Using this information
one can optimize the design and parameters to decrease the error
and to prepare a quantum computer for implementation of
complex quantum algorithms.

\subsection{Decoherence}
The influence of environment can be characterized by the
temperature-dependent spin-lattice relaxation time $T_1$ and
spin-spin relaxation time $T_2$, where $T_2\le T_1$.
The relaxation time $T_2$ defines the maximum number of pulses
and consequently the maximum number $L_{\rm max}$
of qubits in each spin chain. Increasing $L_{\rm max}$
is desirable for increasing the number of qubits in the quantum
register and for better measurement. (Because the total
magnetization of the ensemble of the spin chains is proportional
to $L_{\rm max}$.) We note that the number of the spin chains
$R_{\rm max}$ theoretically is not limited, so that the
size of the whole system and the total number of spins
can be increased by increasing $R$. If $L\ll L_{\rm max}$
the influence of environment is small and one can formulate
quantum dynamics in terms of wave function instead of using
density matrix. This allows one to consider quantum logic operations
with many qubits~\cite{Lopez00,JAM,fullAdder}
and to analytically estimate the influence of other sources
of error which, as shown below, can cause a much more profound
destructive effect on quantum computation.

We define $L_{\rm max}$ from the condition
\begin{equation}
\label{maximumT1}
\tau_{\rm H}+\sum_{j=1}^{L_{\rm max}-1}\tau_j\le T_2,
\end{equation}
where $\tau_j=\pi/\Omega_j$.
The pulse performing the Hadamard transform
can be made very short by increasing the amplitude $B^1_H$
of this pulse, so that we will neglect contribution of
$\tau_H$.
Using Eqs.~(\ref{2piK})-(\ref{maximumT1})
we obtain
\begin{equation}
\label{maximumL}
\sum_{j=1}^{L_{\rm max}-1}
\left(\sum_{l=0}^{j-1}{1 \over (j-l)^3}\right)^{-1}\le
{|J|T_2\over \pi A^3\sqrt{4K^2-1}}.
\end{equation}
The left-hand side for $L_{\rm max}\gg 1$ can be approximated as
\begin{equation}
\label{zeta_approximation}
\sum_{j=1}^{L_{\rm max}-1}
\left(\sum_{l=0}^{j-1}{1 \over (j-l)^3}\right)^{-1}\approx
{L_{\rm max}-1\over\zeta(3)}+0.3399,
\end{equation}
where $\zeta(3)\approx 1.202057$, and $\zeta(x)$ is the Riemann zeta
function.

We now estimate $L_{\rm max}$ for endohedral fullerenes. The very
sharp electron spin resonance spectra from the endohedrals
$^{15}$N@C$_{60}$ and $^{31}$P@C$_{60}$ indicates very long
longitudinal relaxation time $T_1\sim 10$ ms at the temperature
${\cal T}=10$ K, and $T_1\sim 0.9$ ms at ${\cal T}=300$ K. The
transverse relaxation time is $T_2\sim 20~\mu$s at ${\cal T}=50$ K
and $T_2\sim 13~\mu$s at ${\cal T}=300$ K~\cite{fullerenT}. No
nuclear relaxation times have yet been recorded but they are
expected to be several orders of magnitude longer than the
electronic relaxation times. We will make our estimation of $L_{\rm
max}$ for the electron spins. The value of the coupling constant
$|J_e|$ for two electron spins is ($\mu_{\rm B}=9.274\times
10^{-21}$ erg/G) $|J_e|/(2\pi)\approx 52$ MHz. The minimum distance
between two endohedrals is close to the diameter of the C$_{60}$
cage (1 nm) and can be as small as 1.1 nm~\cite{fulleren03}. Taking
$K=1$, $T_2=20~\mu$s, and $A=2.2$, the right-hand side of
Eq.~(\ref{maximumL}) becomes equal to 113. From
Eq.~(\ref{zeta_approximation}) this value corresponds to $L_{\rm
max}=136$. The value of $L_{\rm max}$ decreases when the distance
$A$ between neighboring endohedrals increases.

Due to the interaction of the spin system with
the environment $M$ becomes positive.
The reason is that the energy of the state $|1\rangle$ is larger
than the energy of the state $|0\rangle$,
so that the influence of the environment causes
the transitions $|1\rangle\rightarrow|0\rangle$ while
the transitions $|0\rangle\rightarrow|1\rangle$ are suppressed.

\subsection{Nonresonant transitions}
\begin{figure}
\vspace{-10mm}
\includegraphics[width=8.5cm,height=8cm]{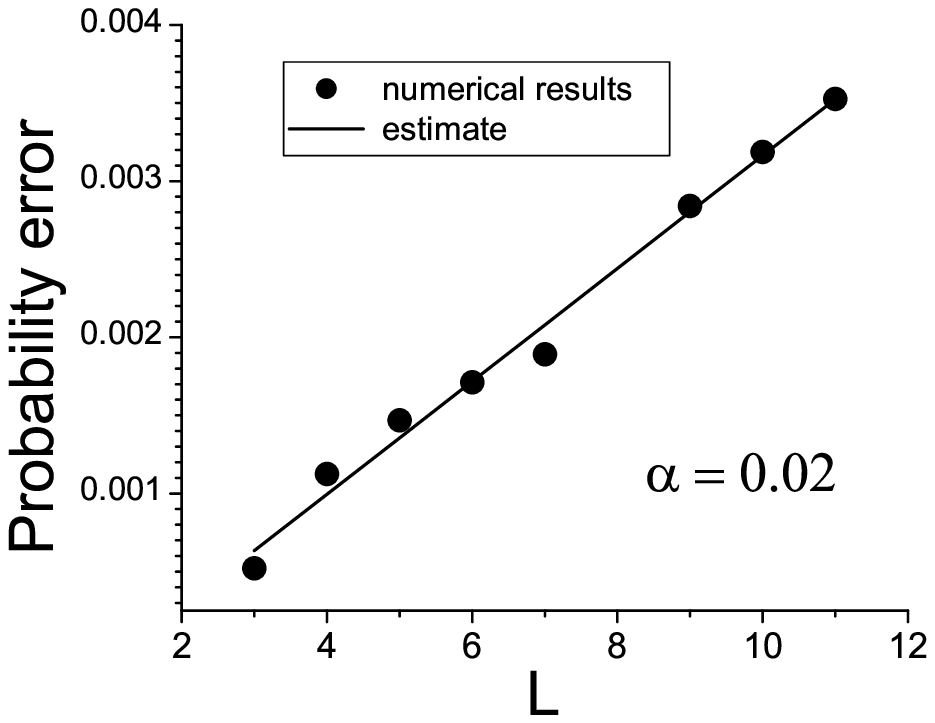}\hspace{-9mm}
\includegraphics[width=8.5cm,height=8cm]{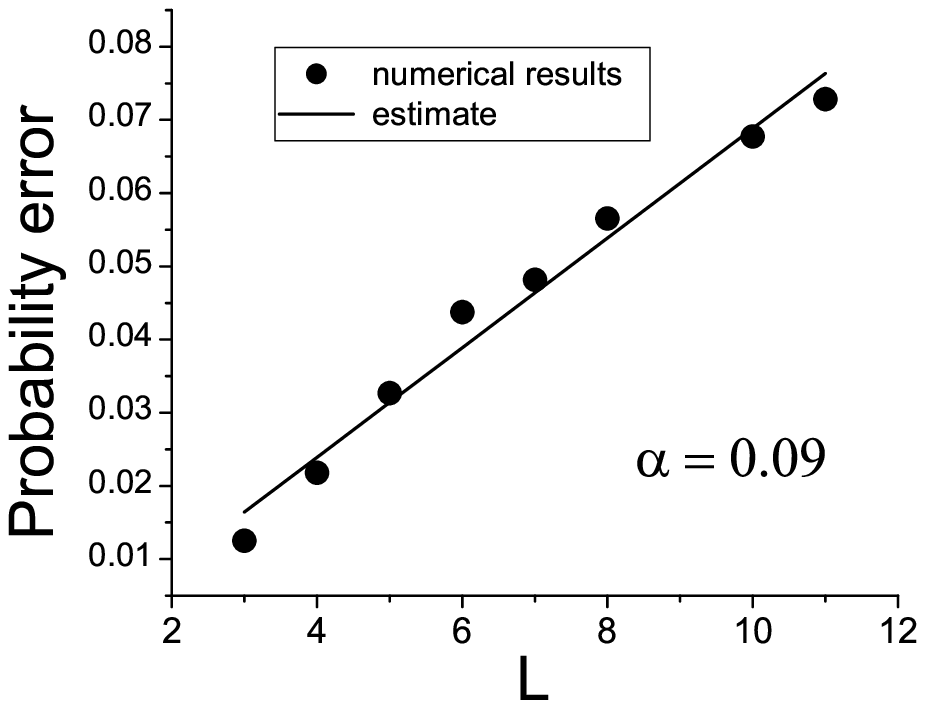}
\vspace{-10mm} \caption{The probability error $P_{\rm nr}$ obtained
using numerical solution~\cite{JAM} and estimate using
Eqs.~(\ref{P_nr}) and (\ref{P_nr01}) as a function of the number of
qubits $L$ for two values of $\alpha$.} \label{reffig:2}
\end{figure}

Since wavelength of the radio-frequency pulses is much larger than
the size of the quantum register, the pulses affect all spins.
If the pulse frequency is close to the frequency $\omega_j$
of $j$th spin then the probability of flipping $k$th spin,
$k\ne j$ is of order of~\cite{PRA01}
$$
\epsilon_{jk}=\left({\Omega_j\over 2|j-k|\delta\omega}\right)^2.
$$
The probabilities of the nonresonant transitions are small
provided that the ratio $\Omega_j/(2|\delta\omega|)$ is small.
Using Eqs.~(\ref{2piK}) and (\ref{detuning}) we obtain
\begin{equation}
\label{alpha}
\epsilon_{jk}\sim{\alpha^2\over 4(j-k)^2},~~~
\alpha={|J|\over \sqrt{4K^2-1}A^3|\delta\omega|}.
\end{equation}
The probabilities of unwanted quantum states created in the result
of the nonresonant transitions are of the order of $\alpha^2$, and
the probability error $P_{\rm nr}$ caused by the nonselective
excitations is proportional to $\alpha^2$. The error $P_{\rm nr}$
grows linearly with the number of pulses $L$~\cite{fullAdder}. A
typical behavior of $P_{\rm nr}(L)$ is shown in Fig.~\ref{reffig:2}
for a small number of qubits $L$ and for two values of $\alpha$:
$\alpha=0.02$ and $\alpha=0.09$. For example, for two electron spins
with $|J|/(2\pi)=52$ MHz separated by the distance 2.2 nm and for
K=1 these values of $\alpha$ correspond, respectively, to
$|\delta\omega|/(2\pi)=141$ MHz and $|\delta\omega|/(2\pi)=31.4$ MHz.
These values of $\delta\omega$ correspond, respectively, to the
following magnetic field gradients ($\delta\omega=\gamma_e\delta
B^0$, $\gamma_e/(2\pi)\approx 28.025$ GHz/T): $2.3\times 10^6$ T/m
and $5.1\times 10^5$ T/m, which can be realized
experimentally~\cite{dH1,dH2,dH3,dH4}.

\begin{figure}
\vspace{-5mm}
\includegraphics[width=8.5cm,height=8cm]{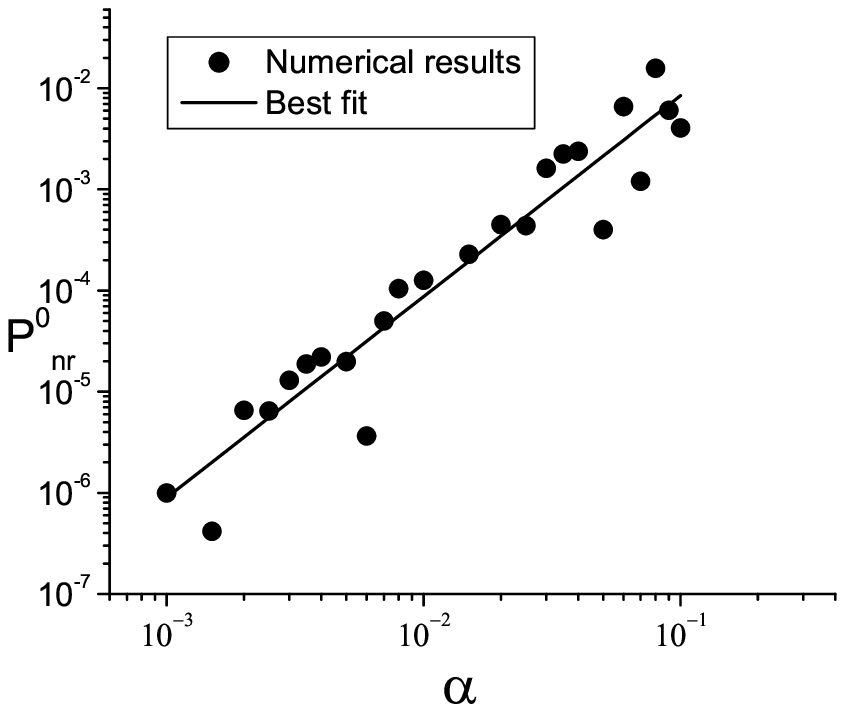}\hspace{-9mm}
\includegraphics[width=8.5cm,height=8cm]{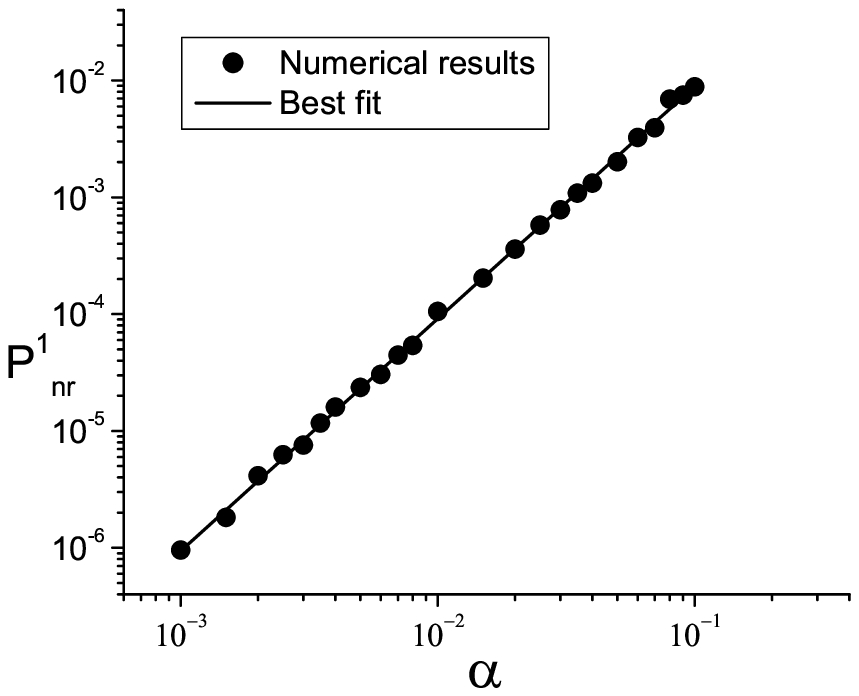}
\vspace{-10mm}
\caption{
$P_{\rm nr}^0$ and  $P_{\rm nr}^1$
in Eq.~(\ref{P_nr}) obtained using numerical
solution and the best fits with the parameters defined by
Eq.~(\ref{P_nr01}).}
\label{reffig:3}
\end{figure}

We approximate $P_{\rm nr}$ as
\begin{equation}
\label{P_nr}
P_{\rm nr}(L)=-P_{\rm nr}^0+P_{\rm nr}^1 L,~~~L>2.
\end{equation}
The values of $P_{\rm nr}^0$ and $P_{\rm nr}^1$ found numerically
are shown in Fig.~\ref{reffig:3}. The numerical simulations are
performed by diagonalization of the full Hamiltonian $H_n$ given by
Eq.~(\ref{Hn}) for each $n$th pulse in the rotating frame where
$H_n$ is time-independent. The obtained eigenstates were used for
simulation of the quantum dynamics~\cite{book1}. The best fit
obtained from the data presented in Fig.~\ref{reffig:3} gives us the
expressions
\begin{equation}
\label{P_nr01}
P_{\rm nr}^0=0.8236\alpha^{1.988},~~~
P_{\rm nr}^1=0.8615\alpha^{1.987},~~~\alpha\ll 1
\end{equation}
In Fig.~\ref{reffig:2} we use these parameters
to approximate the function $P_{\rm nr}(L)$.
One can see that there is a good correspondence between
the numerical results and our estimate~(\ref{P_nr}).
Our estimate~(\ref{P_nr}) is especially useful when $L$ is large,
$L\sim 10^2-10^3$, when no exact solution is available.

\begin{figure}
\vspace{-7mm}
\includegraphics[width=8.5cm,height=8cm]{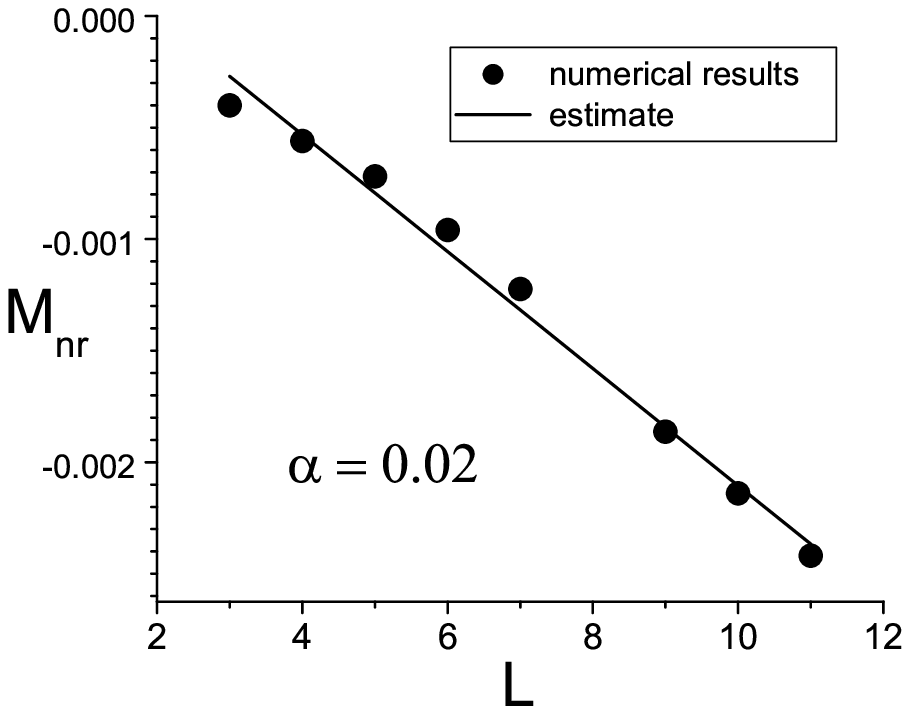}\hspace{-9mm}
\includegraphics[width=8.5cm,height=8cm]{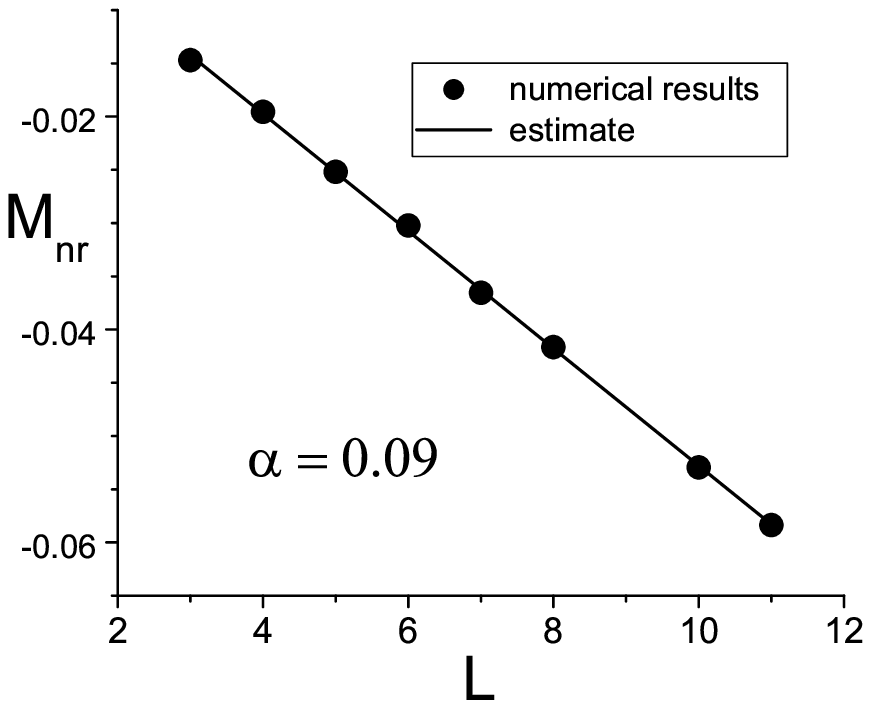}
\vspace{-10mm}
\caption{The dimensionless $z$ component
$M_{\rm nr}$ of the magnetization generated in result of the nonresonant
transitions as a function of the number of qubits $L$ for two
values of $\alpha$. The estimates are calculated using
Eq.~(\ref{M_alpha_beta}) with the parameters defined by
Eq.~(\ref{beta_b}).}
\label{reffig:4}
\end{figure}

Since the z component of the magnetization $M_{\rm nr}$ is proportional
to $P_{\rm nr}$ it is reasonable to present
the dependence of $M_{\rm nr}$ on $L$ in the form
\begin{equation}
\label{M_alpha_beta}
M_{\rm nr}=M_{\rm nr}^0-M_{\rm nr}^1L.
\end{equation}
A typical behavior of $M_{\rm nr}$ as a function of $L$ is shown in
Fig.~\ref{reffig:4} for two values of $\alpha$. From this figure one
can see that $M_{\rm nr}$ is negative.

\begin{figure}
\vspace{-5mm}
\includegraphics[width=8.5cm,height=8cm]{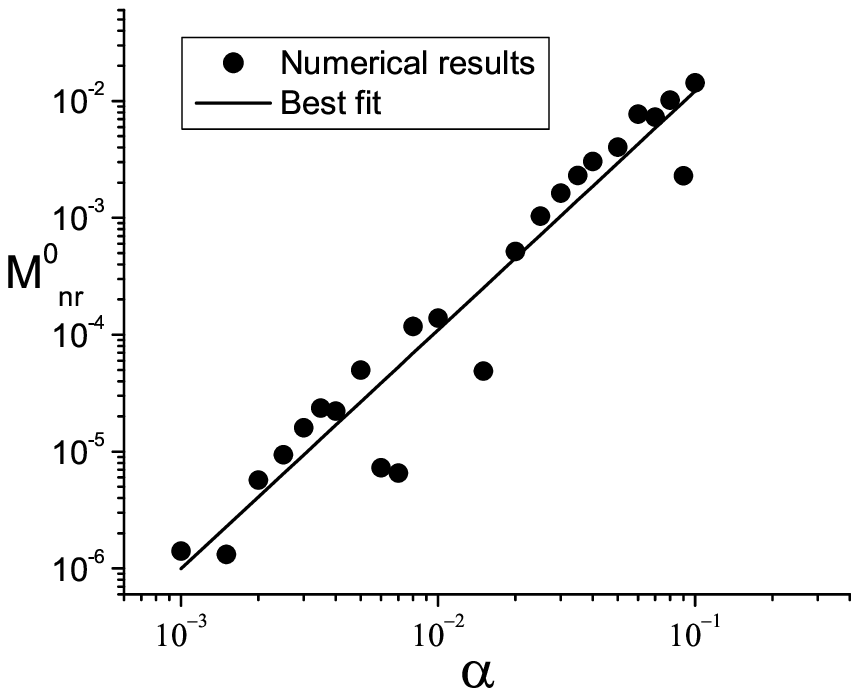}\hspace{-9mm}
\includegraphics[width=8.5cm,height=8cm]{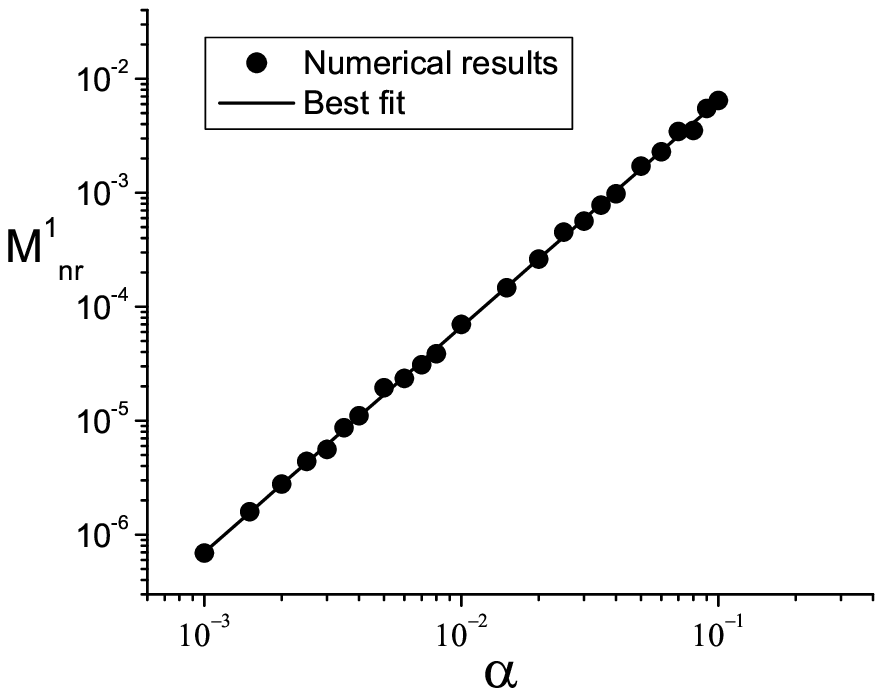}
\vspace{-10mm}
\caption{$M_{\rm nr}^0$ and $M_{\rm nr}^1$ as a function of $\alpha$.
The best fit gives the parameters defined by Eq.~(\ref{beta_b}).}
\label{reffig:5}
\end{figure}

From Fig.~\ref{reffig:4} one can numerically calculate $M_{\rm
nr}^0$ and $M_{\rm nr}^1$. We calculated $M_0$ and $M_1$ for
different values of $\alpha$ (see Fig.~\ref{reffig:5}) and obtained
\begin{equation}
\label{beta_b}
M_{\rm nr}^0=1.341\alpha^{2.044},~~~
M_{\rm nr}^1=0.60786\alpha^{1.9795},~~~\alpha\ll 1.
\end{equation}
From Fig.~\ref{reffig:3} one can see that Eq.~(\ref{M_alpha_beta})
with parameters (\ref{beta_b}) gives us a good approximation of
$M_{\rm nr}$. It is important that the number of qubits in
Eqs.~(\ref{P_nr}) and (\ref{M_alpha_beta}) is an explicit
parameter, so that one can calculate the probability errors
and $z$ component of the magnetization due
to the nonresonant transitions for an arbitrary number of qubits $L$.
Combining Eqs.~(\ref{P_nr}) and (\ref{M_alpha_beta}) we obtain
the relation between $M_{\rm nr}$ and $P_{\rm nr}$
\begin{equation}
\label{MP}
M_{\rm nr}=g^0_{\rm nr}+g^1_{\rm nr}P_{\rm nr},
\end{equation}
where
$$
g^0_{\rm nr}=M^0_{\rm nr}-{P^0_{\rm nr}\over P^1_{\rm nr}}M^1_{\rm nr},~~~
g^1_{\rm nr}=-{M^1_{\rm nr}\over P^1_{\rm nr}}.
$$

\subsection{Interaction between different chains}
\label{sec:differentChains} 
\begin{figure}[t!h!b!]
\vspace{-5mm}
\centerline{\includegraphics[width=12cm,height=8cm,angle=0,clip=]{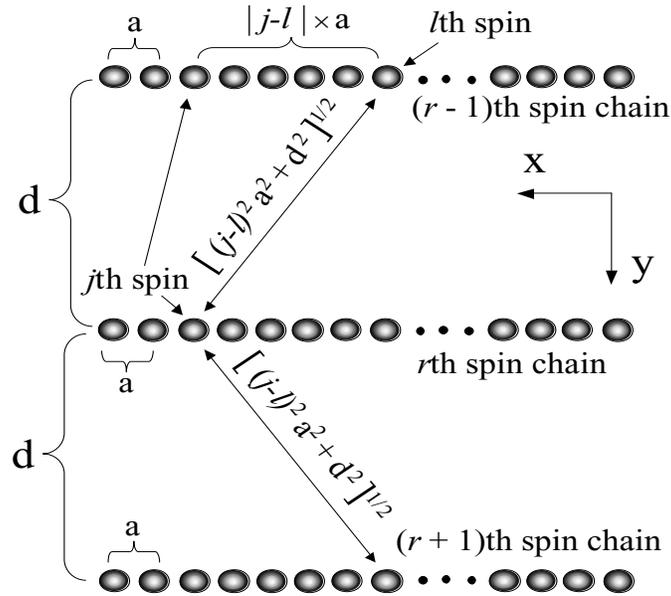}}
\vspace{-2mm}
\caption{A scheme for calculation of the influence
of the $l$th spin of the $(k-1)$th and $(k+1)$th spin chains on the
$j$th spin of the $k$th spin chain.}
\label{reffig:chains}
\end{figure}

Here we will discuss the possibility to
decrease the influence of the chains on each other by optimal
choice of the parameters of the pulses. We will show that if
the long-range interaction between the chains
is not taken into consideration the influence of the
chains on each other causes the error of the order of $L(a/d)^3$.
Consider the $j$th spin of the $r$th spin chain in the field of
the spins of the $(r-1)$th, $r$th and $(r+1)$th chains. (See
Fig.~\ref{reffig:chains}.) The resonant frequency is [compare with
Eq.~(\ref{frequency})]
\begin{equation}
\label{frequency1}
\nu_j=\omega_j+
{J\over A^3}\sum_{l=0\atop l\ne j}^{L-1}{s^z_l(p^\prime)\over |l-j|^3}
+{2J\over D^3}\sum_{l=0\atop l\ne j}^{L-1}
{s^z_l(p^\prime)\over[1+\chi^2(j-l)^2]^{3/2}},
\end{equation}
where $D$ is the dimensionless
distance between neighboring chains measured in nanometers and
$\chi=a/d=A/D\ll 1$. The Rabi frequency of the pulse is
[see Eqs.~(\ref{2piK}) and (\ref{detuning})]
\begin{equation}
\label{detuning1}
\Omega_j={|J|\over A^3\sqrt{4K^2-1}}
\sum_{l=0}^{j-1}{1\over (j-l)^3}
+{2|J|\over D^3\sqrt{4K^2-1}}
\sum_{l=0}^{j-1}{1\over [1+\chi^2(j-l)^2]^{3/2}}.
\end{equation}

We now can estimate the error caused by the influence of different
chains on each other if we neglect the interaction between them.
Consider the resonant transition for the excited state of the superposition
using Eq.~(\ref{2x2}). If we disregard the influence of the
neighboring chains on each other, then instead of the resonant
transition with $\Delta_j(q^\prime,p^\prime)=0$ we have the nonzero
contribution of the third term in Eq.~(\ref{frequency1})
$$
|\Delta_j(q^\prime,p^\prime)|=
{2|J|\over D^3}\left|\sum_{l=0\atop l\ne j}^{L-1}
{s^z_l(p^\prime)\over [1+\chi^2(j-l)^2]^{3/2}}\right|.
$$
Since the value of the sine in the first Eq.~(\ref{2x2}) is of the
order of unity and the value of the cosine is close to zero, the
value of the coefficient $C_P(t_{j}+\tau_j)$ is of the order
of (instead of zero in the ideal case)
$$
|C_{p^\prime}(t_{j}+\tau_j)|\approx
\left|{\Delta_j(q^\prime,p^\prime)\over\lambda_j(q^\prime,p^\prime)}
\right|\approx
\left|{\Delta_j(q^\prime,p^\prime)\over\Omega_j}\right|
$$
\begin{equation}
\label{chainChain}
\approx 2\sqrt{4K^2-1}\chi^3\left|\sum_{l=0\atop l\ne j}^{L-1}
{s^z_l(p^\prime)\over [1+\chi^2(j-l)^2]^{3/2}}\right|
\left[\sum_{l=0}^{j-1}{1\over (j-l)^3}\right]^{-1}.
\end{equation}
The error $P_{\rm int}$ caused by the influence of neighboring
chains on each other is of the order of $P_{\rm
int}=|C_{p^\prime}(t_{j}+\tau_j)|^2$. From Eq.~(\ref{chainChain})
one can see that $P_{\rm int}$ increases with the Rabi frequency
$\Omega_j$ decreasing.

In order to estimate the error in the probability amplitude
given by Eq.~(\ref{chainChain}) consider a typical example.
Let the distance between the chains be equal to the size of a
single chain, $d=(L-1)a$ [$D=(L-1)A$], $L\gg 1$. Then $\chi=1/(L-1)$.
Let us estimate $\sqrt{P_{\rm int}}$ for $j=L-1$.
The first sum in Eq.~(\ref{chainChain}) is 
$$
-{1\over 2}\sum_{l=0}^{L-2}
{1\over [1+\chi^2(L-1-l)^2]^{3/2}}=
-{1\over 2}\sum_{l=0}^{L-2}
{\left[1+\left(1-{l\over L-1}\right)^2\right]^{-3/2}}.
$$
For all terms of the sum we have
$$
{1\over 2^{3/2}}\le\left[1+\left(1-{l\over L-1}\right)^2\right]^{-3/2}
<1,
$$
so that
$$
{L-1\over 2^{3/2}}<\sum_{l=0}^{L-2}
{\left[1+\left(1-{l\over L-1}\right)^2\right]^{-3/2}}<L-1.
$$
For the second sum we have
$$
\sum_{l=0}^{L-2}{1\over (L-1-l)^3}=1+{1\over 2^3}+
{1\over 3^3}+\dots+{1\over(L-1)^3}<\zeta(3)\approx 1.202.
$$
Finally, assuming $K=1$ we obtain
\begin{equation}
\label{sqrtPint}
\sqrt{P_{\rm int}}>0.51(L-1)\chi^3.
\end{equation}
This is the error introduced by only one pulse.
The errors generated by different pulses of the protocol can
accumulate.

From Eq.~(\ref{sqrtPint}) one can see that a small parameter which
characterizes the dipole-dipole interaction between the chains is
$L\chi^3$ rather than $\chi^3$. The influence of neighboring chains
on each other can be minimized by introducing the corrections to the
frequency [the third term in the right-hand side of
Eq.~(\ref{frequency1})] and to the Rabi frequency [the second term
in the right-hand side of Eq.~(\ref{detuning1})]. These corrections
minimize the errors only for intermediate chains and do not minimize
the error for the edge chains with $r=0$ and $r=R-1$. One can use
our corrections if the number of chains $R$ is large, so that one
can neglect the edge chains, or when the chains are placed
relatively close to each other, when these corrections are
relatively large.

If one minimizes the errors caused by the nearest-neighboring
chains, the error in the probability amplitude caused by the
influence of the next-nearest neighboring chains is of the order of
$\sqrt{P_{\rm int}}/8$. Using our approach one can minimize the errors
caused by the next-nearest neighboring chains and disregard
the chains with $r=0,1,R-2,R-1$. This makes the error caused
by the next-next-nearest neighboring chains to be of the order
of $\sqrt{P_{\rm int}}/27$ and so on.

\subsection{Qubit displacements}
One of the most serious problems that prevents building a
solid-state quantum computer is manufacturing the spin system like
that shown in Fig.~1. Atoms with nonzero spin, such as $^{31}$P, can
be placed on the surface of a magnetically neutral substance, such
as $^{28}$Si, using, for example, scanning tunneling microscopy
technique~\cite{Hawley01,Tucker}. The placement of the qubits can be
not perfect, so that these qubits form distorted spin chains. If one
deals with a single chain, one can measure the  locations of the
qubits and to chose the suitable pulse parameters to compensate the
deviations of the qubits from their prescribed positions. On the
other hand, if one implements a quantum algorithm on an ensemble of
spin chains, the deviation in the location of a qubit from the
perfect position (called below displacement) in a chain makes this
chain different from other chains, and this error cannot be
completely compensated by a proper choice of the parameters of the
pulses. Here we will investigate this kind of error.

Since a qubit  in a solid state is usually incorporated
into the crystal lattice, the minimum possible
qubit displacement is equal to the lattice constant.
If the displacement happens in the direction of the magnetic
field gradient (along the $x$ axis in
Figs.~1 and \ref{reffig:chains}), then even a
small displacement causes a relatively large change in the
Larmor frequency of this qubit because the magnetic field
gradient is supposed to be large. Consequently, we believe
that this kind of error causes the most profound
destructive effect on quantum computation in our system.
Since the frequency of the displaced $k$th qubit considerably differs
from the frequency of the pulse, this qubit will not flip and
the other $(k+1)$th, $(k+2)$th, \dots, $(L-1)$th qubits will not
flip also. For example, if $k=3$ and $L=6$ the excited state evolves as
$$
|0_50_4{\bf 0}_30_2\underline{0}_11_0\rangle\rightarrow
|00{\bf 0}\underline{0}11\rangle\rightarrow
|000\underline{{\bf 0}}111\rangle\rightarrow
|0\underline{0}{\bf 0}111\rangle\rightarrow
|\underline{0}0{\bf 0}111\rangle\rightarrow|00{\bf 0}111\rangle,
$$
where the qubit to be flipped by the corresponding pulse is underlined.
One can see that the $z$ component $M$ of
the magnetization due to the error caused by displacement of qubits
is positive.

We now calculate the error due to the qubit displacement(s).
It is convenient to define the dimensionless displacement $v_k$
of the $k$th qubit as
\begin{equation}
\label{v}
v_k={|d\omega_k|\over\delta\omega}=\pm {|dx_k|\over a},
\end{equation}
where $dx_k$ is the dimensional displacement,
$d\omega_k$ is the change of the Larmor frequency of the $k$th
qubit caused by this displacement, the sign ``+'' must be used if 
one considers nuclear spins and the sign ``-'' must be used for electron spins. 
For example, the value
$v_k=1/15$ corresponds to the displacement by one lattice site
if the number of atoms between neighboring qubits is equal to 14
and by two lattice sites if the number of atoms between neighboring
qubits is equal to 29.

If the probability ${\cal P}$ of a displacement is relatively large,
${\cal P}\ge 1/L$ then the number of ``perfect'' spin chains,
where all spins are not displaced, is relatively small,
so that it is important first to study the errors $P_{\rm d}$ and the
magnetization $M_{\rm d}$ caused by the qubit displacements
in a single spin chain.

The displaced $k$th qubit affects all other qubits in the chain.
The transition frequency and detuning of the $j$th qubit ($j\ne k$)
change by the value [see Eq.~(\ref{frequency})]
\begin{equation}
\label{alphaDisplacement}
|\Delta_{jk}|={|J|\over 2A^3}\left|{1\over |k\pm v_k-j|^3}-{1\over |k-j|^3}\right|
\approx {3|J|\over 2A^3} {|v_k|\over|k-j|^4}\ll \Omega_j.
\end{equation}
It is convenient to characterize the influence of the displaced qubit
on all other qubits by a small dimensionless parameter
\begin{equation}
\label{beta}
\beta_{jk}=
{3v_k\sqrt{4K^2-1}\over 2(j-k)^4}\left[\sum_{l=0}^{j-1}{1\over (j-l)^3}\right]^{-1}
\approx{|\Delta_{jk}|\over\Omega_j}\ll 1.
\end{equation}
Then in Eq.~(\ref{2x2}) we have
$$
\sin\left({\lambda_j\tau\over 2} \right) \approx 1-{\pi^2\over
32}\beta_{jk}^4\approx 1,~~~ \cos\left({\lambda_j\tau\over 2}
\right)\approx -{\pi\over 2}\beta_{jk}^2.
$$
When the $j$th qubit is flipped in the excited state we have
$\Delta=\Delta_{jk}$ instead of $\Delta=0$ and the error is
$$
P_{jk}\approx{\Delta_{jk}^2\over\lambda_{jk}^2}=
{1\over 2}\beta_{jk}^2\approx{9(4K^2-1)\over 8(j-k)^8\zeta^2(3)}v_k^2.
$$
For example, for $v_k=1/20$ and $K=1$ we have
$$
P_{k\pm1,k}\approx 0.006,~~~P_{k\pm 2,k}\approx 2.3\times 10^{-5}.
$$
The probability of the transition from the ground state caused by
the deviation in the detuning is
$$
P_{jk}^\prime\approx{\pi^2(4K^2-1)\over 128 K^4}\beta_{jk}^2.
$$
For $K=1$ the error caused by the ground state
$P_{jk}^\prime\approx 0.25\beta_{jk}^2$ is of the same order as the
error $P_{jk}$ caused by the excited state. Below we will neglect
the errors $P_{jk}$ and $P_{jk}^\prime$ caused by the influence
of the displaced qubit on all other qubits as being small compared to the
other errors.

We now estimate the probability $P_{kk}$ caused by the $k$th pulse
on the $k$th displaced qubit.
Since $\alpha\sim 1/|\delta\omega|$ it is convenient to measure the
dimensionless frequency displacement in units of $1/\alpha$, so that
the change in the transition frequency of the $k$th qubit caused by its
displacement is proportional to $v_k/\alpha$. We first analyze the action
of the $k$th pulse on the excited state.
Instead of the resonant transition with the detuning
$\Delta_k=0$ we have the transition with the detuning
$\Delta_k=\pm v_k\delta\omega$,
where the sign ``+'' corresponds to the displacement in the positive
$x$ direction and the sign ``-'' correspond to the displacement in the
opposite direction.

If  $k\ne 0$ then the amplitude of the excited state is
$$
|C_{q^\prime}|={1\over \sqrt 2}
{1\over \sqrt{1+\left(v_k/\alpha_k\right)^2}}
\sin\left[{\pi\over 2}\sqrt{1+\left({v_k\over \alpha_k}\right)^2}\right],
$$
where
\begin{equation}
\label{alpha_k}
\alpha_k={\Omega_k\over |\delta\omega|}=
\alpha\sum_{l=0}^{k-1}{1\over |k-l|^3}.
\end{equation}
The error generated by the excited state is
\begin{equation}
\label{Pd}
P_{kk}\equiv P_{\rm d}=
{1\over 2}\left\{1-{1\over 1+\left(v_k/\alpha_k\right)^2}
\sin^2\left[{\pi\over 2}\sqrt{1+\left({v_k\over \alpha_k}\right)^2}
\right]\right\}.
\end{equation}
One can see that the ratio $v_k/\alpha$ characterizes the
error caused by the displacement $v_k$: if
$v_k/\alpha\rightarrow 0$ then $P_{\rm d}\rightarrow 0$, otherwise
$P_{\rm d}\rightarrow 1/2$.

The error caused by the action of the $k$th pulse, $k\ne 0$, on the
ground state of the superposition is
\begin{equation}
\label{Pd^prime}
P_{\rm d}^\prime(\pm)=
{1\over 2\left[1+\left(\sqrt{4K^2-1}\mp{v_k\over \alpha_k}\right)^2\right]}
\sin^2\left[{\pi\over 2}
\sqrt{1+\left(\sqrt{4K^2-1}\mp{v_k\over \alpha_k}\right)^2}\right].
\end{equation}
For a large magnetic field gradient the error
$P_{\rm d}^\prime$ is small,
\begin{equation}
\label{Pdprime_small} P_{\rm d}^\prime\sim {1\over
\left(v_k/\alpha_k\right)^2},~~~{|v_k|\over \alpha_k}\gg 1.
\end{equation}

For $k=0$ the probability of the excited state after implementation
of the Hadamard transform on the displaced qubit is
\begin{equation}
\label{calP}
\eta=|C_1|^2={1\over 1+\left(v_0/\alpha_0\right)^2}
\sin^2\left[{\pi\over 4}\sqrt{1+\left({v_0\over \alpha_0}\right)^2}
\right],
\end{equation}
where $\alpha_0=\Omega_{H}/|\delta\omega|$,
$\Omega_{H}$ is the Rabi frequency of the pulse
implementing the Hadamard transform. Here and in the
sequel  we take $\alpha_0=\alpha_1=\alpha$.
The probability of error generated by this pulse is
$$
\left|{1\over 2}-|C_0|^2\right|+\left|{1\over 2}-|C_1|^2\right|,
$$
where $|C_0|^2=1-|C_1|^2$.
Assuming that the other pulses of the protocol do not generate error,
we obtain that the error due to the displaced zeroth qubit is
\begin{equation}
\label{Pdprime}
P_{00}\equiv P_{\rm d}^{\prime\prime}=
\left|{1\over 2}-(1-\eta)\right|+\left|{1\over 2}-\eta\right|=1-2\eta.
\end{equation}

\begin{figure}
\vspace{-10mm}
\centerline{\includegraphics[width=14cm,height=10cm]{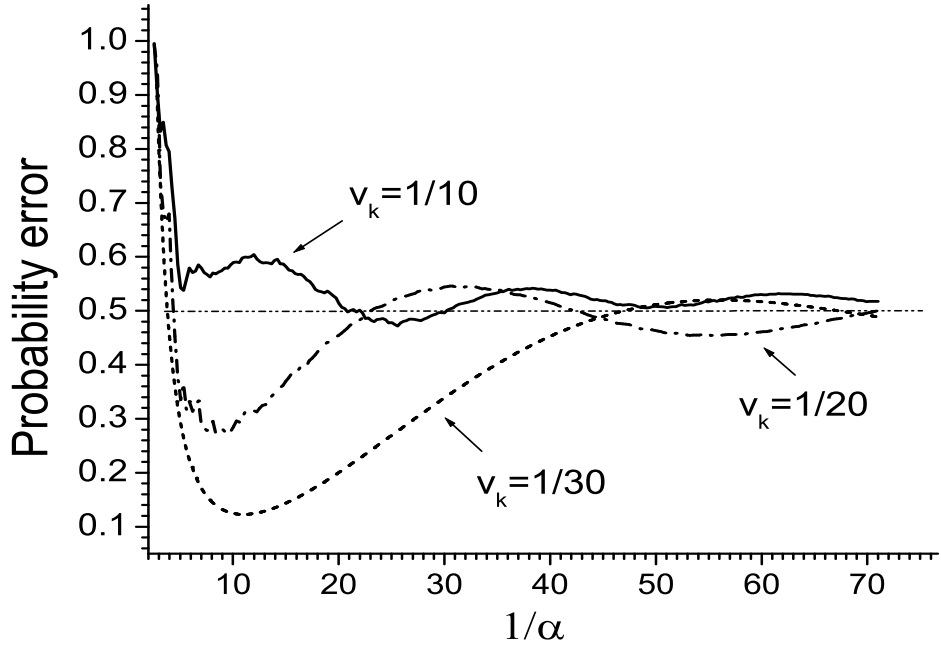}}
\vspace{-10mm}
\caption{The probability error $P$ obtained using
numerical solution as a function of $1/\alpha$
for different values of $v_k$. The displaced qubit is located
at the center of the chain, $L=9$.}
\label{reffig:7}
\end{figure}

Since $1/\alpha\sim|\delta\omega|$ decreasing $|\delta\omega|$
decrease the error $P_{\rm d}$. On the other hand, due to
Eq.~(\ref{P_nr}) and (\ref{P_nr01}), the error $P_{\rm nr}$
caused by the nonselective excitations (nonresonant transitions)
increases with $|\delta\omega|$ decreasing.
The total probability error for $k\ne 0$ is
\begin{equation}
\label{Ptotal}
P=P_{\rm nr}+P_{\rm d}+P_{\rm d}^\prime(-),
\end{equation}
where the qubit is assumed to be displaced in the negative $x$ direction.
In Fig.~\ref{reffig:7} we plot the probability error $P$, defined as
\begin{equation}
\label{errorDefinition}
P=\left|{1\over 2}-|C_0(T)|^2\right|+\left|{1\over 2}-|C_{2^L-1}(T)|^2\right|,
\end{equation}
which was obtained using exact numerical solution for $k\ne 0$.
In Eq.~(\ref{errorDefinition}) $T$ is the total time of implementation of
the entanglement protocol. When $1/\alpha$ is small
the probability error $P$ is large due to the nonresonant excitations.
When $1/\alpha$ is large $P$ is large because the displaced qubit
does not flip. From the results presented in Fig.~\ref{reffig:7}
one can see that if the displacement is relatively large
($v_k=1/10$), then the error is always large
and the entanglement protocol cannot be optimized for any
parameters of the model.

As follows from Fig.~\ref{reffig:7}, the magnitude of $P$ becomes
close to 1/2 and relatively independent of $L$ and $1/\alpha$ when
$1/\alpha$ is large, i.e. when $(v_k/\alpha)^2\gg 1$. For these parameters
we have mostly two states in the superposition: the ground state and
the partially excited state. For example, if the $k$th qubit is displaced, $k=3$
and $L=6$, then instead of the desired entangled state
\begin{equation}
\label{entangledL6}
{1\over \sqrt 2}\left(|00000\rangle+|111111\rangle\right)
\end{equation}
we have the state
$$
{1\over \sqrt 2}\left(|00000\rangle+|000111\rangle\right).
$$

The $z$ component of the magnetization $M_{\rm d}$ due to a displaced qubit
can be estimated using the probability error.
If $k\ne 0$ the probability $P_{\rm d}$ is mostly independent of the
position $k$ of the displaced qubit in the chain.
Unlike $P$, $M_{\rm d}$ is large (and positive) if
the displaced qubit is located in the beginning of the spin chain
and relatively small if the displaced qubit is located in the end of the chain.
For example,
if $k$ is the number of the displaced qubit and $k=1$ then the excited state
is ($L=6$) $|000001\rangle$ instead of $|111111\rangle$ and the entangled
state is
\begin{equation}
\label{displaceM}
{1\over \sqrt 2}\left(|00000\rangle+|000001\rangle\right)
\end{equation}
instead of the state~(\ref{entangledL6}).
The $z$ component of the magnetization for to the state (\ref{displaceM}) is
$M_{\rm d}=(5/6)$. If the displaced qubit is located in the end
of the chain, for example, if $k=6$, then the value of $M$ for the state
\begin{equation}
\label{displaceM1}
{1\over \sqrt 2}\left(|00000\rangle+|011111\rangle\right)
\end{equation}
is $M_{\rm d}=(1/6)$.

In order to relate $M$ with the probability error consider the
two situations when  $k\ne 0$ and when $k=0$. If $k\ne 0$ we note that
after implementation of the entanglement protocol on the spin chain
with a displaced qubit in the register there are mostly four
quantum states: the ground state with the probability
$1/2-P_{\rm d}^\prime(\pm)$,  the error state with the probability
$P_{\rm d}^\prime(\pm)$ created from the ground state,
the error state with the probability $P_{\rm d}$
created from the excited state,
and the fully excited state $|11\dots11\rangle$ with the probability
$1/2-P_{\rm d}$. Next, we assume that the position of the displaced qubit
in the chain is random. By averaging over many random realizations we
obtain that the two error states do not contribute to $M$.
For example, the $z$ component of the magnetization of the partially
excited state in Eq.~(\ref{displaceM}) is
$M={1\over 2}\cdot{4\over 6}$ while
in Eq.~(\ref{displaceM1}) $M=-{1\over 2}\cdot{4\over 6}$, so that
the average of these two contributions is zero. The contribution to the
$M$ due to the fully excited state
is $-\left({1\over 2}-P_{\rm d}\right)$. By adding all these contributions we
obtain
\begin{equation}
\label{MdPd}
M_{\rm d}={1\over 2}-P_{\rm d}^\prime(\pm)-
\left({1\over 2}-P_{\rm d}\right)=P_{\rm d}-P_{\rm d}^\prime(\pm).
\end{equation}
If $k=0$ there are mostly two states in the register: the ground state
with the probability $1-\eta$ and the fully excited state with the probability
$\eta$, where $\eta$ is given by Eq.~(\ref{calP}). The magnetization
due to the displaced zeroth qubit is
\begin{equation}
\label{MdPd1}
M_{\rm d}^{\prime\prime}=1-2\eta=P_{\rm d}^{\prime\prime}.
\end{equation}

The total magnetization is
$$
M=M_{\rm nr}+M_{\rm d},~~{\rm for}~~k\ne 0,
$$
\begin{equation}
\label{Mtotal}
M=M_{\rm nr}+M_{\rm d}^{\prime\prime},~~{\rm for}~~k=0,
\end{equation}
where $M_{\rm nr}$ is given by Eqs.~(\ref{M_alpha_beta}) and (\ref{beta_b}).

\begin{figure}
\includegraphics[width=14cm,height=10cm]{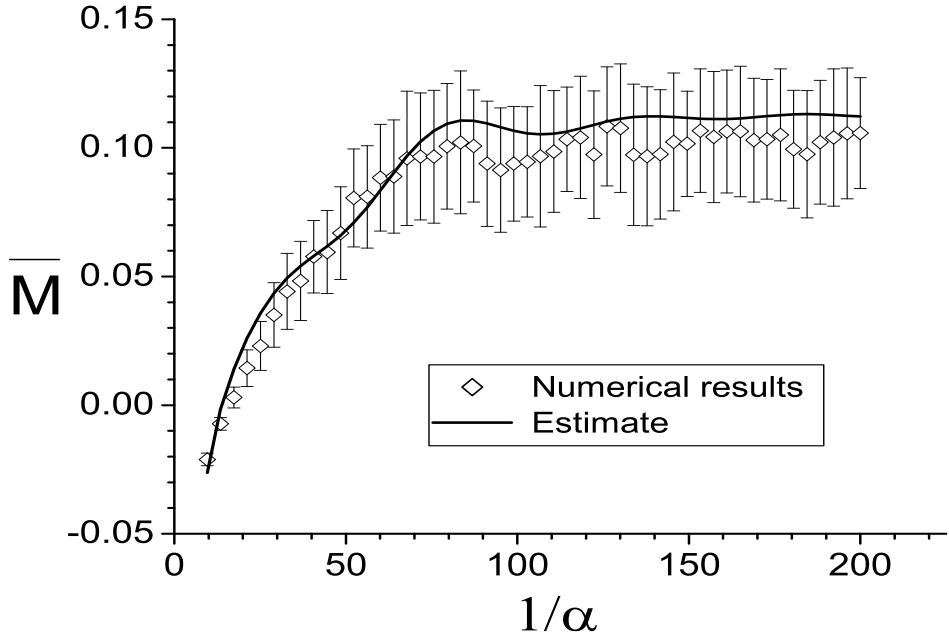}
\vspace{-10mm} \caption{The $z$ component of the magnetization
$\overline M$ as a function of $1/\alpha$ for $R=100$ spin chains
with $L=7$ qubits in each chain. The numerical results are averaged
over 50 realizations of different $7\times 100$ qubit ensembles with
randomly chosen displaced qubits, $\xi=1/(5L)$. The displacements
are in the random directions along the $x$ axis. The estimate is
calculated using Eq.~(\ref{aavMM}). $v_k=1/20$, $K=1$.}
\label{reffig:8}
\end{figure}

Assume that we have an ensemble of $R$ spin chains
and the probability of a qubit to be displaced by one lattice site is $\xi$.
If, for example, $\xi =1/L$, then on average one qubit in each chain
is displaced, if  $\xi =1/(nL)$ then on average one qubit in $n$
chains is displaced. The total
number of displaced qubits with $k\ne 0$ is on average
$\xi R(L-1)$ and the total number of displaced qubits with $k=0$ is
on average $\xi R$, so that
\begin{equation}
\label{Md} \overline{M}_{\rm d}={\xi\over L}
\left[\sum_{k=1}^{L-1}(P_{{\rm d},k}-P_{{\rm d},k}^{\prime})+ P_{\rm
d}^{\prime\prime}\right],~~~ P_{{\rm d},k}^{\prime}\equiv {1\over
2}[P_{{\rm d},k}^{\prime}(-)+P_{{\rm d},k}^{\prime}(+)],
\end{equation}
where $P_{{\rm d},k}$ and $P_{{\rm d},k}^{\prime}$ depend
on $k$ through the dependence
of $\alpha_k$ on $k$ in Eq.~(\ref{alpha_k}). We calculated numerically
\begin{equation}
\label{aavMM}
\overline{M}=M_{\rm nr}+\overline{M}_{\rm d}
\end{equation}
as a quantum-mechanical average for 100 noninteracting spin chains
with randomly chosen displaced qubits and random displacement directions
(in positive and negative directions along the $x$ axis).
In Fig.~\ref{reffig:8} we plot $\overline{M}$ as a function of $\alpha$
obtained using the exact numerical solution and our estimate given
by Eq.~(\ref{aavMM}) for $\xi=1/(5L)$.
As follows from the figure our estimate (\ref{aavMM})
is a good analytical approximation of $M$. In particular,
Eq.~(\ref{aavMM}) can be used for estimation of $M$ when
the number of qubits $L$ in each chain is large.

\subsection{Fluctuations of permanent magnetic field} The error
$P_{\rm osc}$ caused by fluctuations of permanent magnetic field
$d\omega$ due, for example, to unwanted oscillations of the DC
current in wires needed for creating the magnetic field gradient
can be estimated using Eqs.~(\ref{Pd}) and (\ref{Pd^prime}).
Instead of the dimensionless deviation $v_k$ we introduce the
average dimensionless deviation $\bar v$ as
$$
\bar v=\left|{\overline{d\omega}\over \delta\omega}\right|,
$$
where $\left|\overline{d\omega}\right|$
is the average deviation of the transition frequency of a qubit from the optimal
value caused by a fluctuation of the current in the wires
creating the magnetic field gradient. We assume that this deviation
is small, $(\bar v/\alpha)^2\ll 1$.

The average error due to unwanted transitions from the excited
and ground states for $k\ne 0$ is [see Eqs.~(\ref{Pd}) and (\ref{Pd^prime})]
\begin{equation}
\label{Posc}
P_{\rm osc}(k)=P_{\rm osc}^\prime(k)+P_{\rm osc}^{\prime\prime}(k),
\end{equation}
where
$$
P_{\rm osc}^\prime(k)=
{\pi^2(4K^2-1)\over 128K^4}\left({\bar v\over \alpha_k}\right)^2,~~~
P_{\rm osc}^{\prime\prime}(k)={1\over 2}\left({\bar v\over \alpha_k}\right)^2.
$$
Here $P_{\rm osc}^{\prime}(k)$ and $P_{\rm osc}^{\prime\prime}(k)$
are the probability errors created from, respectively, the ground and
excited states by action of the $k$th pulse of the protocol.
For $k=0$ (Hadamard gate)
from Eq.~(\ref{calP}) the probability of the excited state is
$$
\eta\approx {1\over 2}
\left[1+\left({\pi\over 4}-1\right)\left({\bar v\over \alpha_0}\right)^2\right],
$$
so that the average error is
$$
P_{\rm osc}(k=0)=1-2\eta=
\left(1-{\pi\over 4}\right)\left({\bar v\over \alpha_0}\right)^2.
$$
The average total error due to the nonresonant transitions and oscillations
of the magnetic field is
\begin{equation}
\label{P_nr_osc}
P=P_{\rm nr}+P_{\rm osc}(0)+\sum_{k=1}^{L-1}P_{\rm osc}(k)
\end{equation}
Since $P_{\rm nr}\sim\alpha^2$ and $P_{\rm osc}(k)\sim1/\alpha^2$
there is an optimal value of $\alpha$
\begin{equation}
\label{alpha_opt}
\alpha_{\rm opt}\approx\left[{\bar v^2\over (-0.82+0.86L)}\right]^{1/4}
\left\{1-{\pi\over 4}+\left[{1\over 2}+{\pi^2(4K^2-1)\over 128K^4}\right]
\left[{L-1\over \zeta^2(3)}+0.6\right]\right\}^{1/4}
\end{equation}
where $P$ is minimal. For example, for $\bar v=10^{-4}$, $L=9$, and $K=1$
we have $\alpha_{\rm opt}\approx 9.08\times 10^{-3}$.

\begin{figure}
\includegraphics[width=14cm,height=10cm]{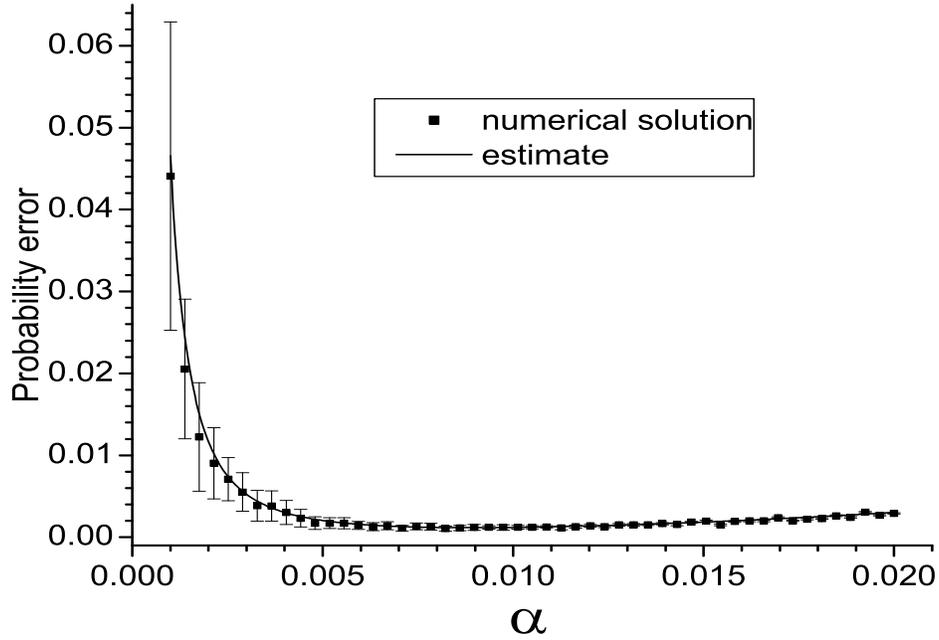}
\vspace{-10mm}
\caption{The error caused by nonresonant transitions and
unwanted oscillations of the permanent magnetic field as a function of
alpha for $\bar v=10^{-4}$ and $L=9$. The estimate is obtained using
Eq.~(\ref{P_nr_osc}).}
\label{reffig:10}
\end{figure}

In Fig.~\ref{reffig:10} we plot the error $P$ caused by action of
both nonresonant transitions and unwanted oscillations of the
permanent magnetic field. Numerical results are obtained by
diagonalization of the full Hamiltonian matrix in the rotating frame.
The random values of $v$ for the numerical results have
the Gaussian distribution with the zero average and dispersion
equal to $\bar v$. In our model a magnetic field fluctuation
is constant (and random) during each pulse.
Each point of the numerical results
is the average over 30 realizations with the random values of $v$.
One can see from the figure that
for $\alpha<\alpha_{\rm opt}$ the error is mostly defined
by the fluctuations of the permanent magnetic field and
$P$ decreases with increasing $\alpha$ as $P\sim 1/\alpha^2$.
For $\alpha>\alpha_{\rm opt}$ the error is due to the nonresonant
excitations and $P\sim \alpha^2$. From Fig.~\ref{reffig:10}
one can see that our estimate (\ref{P_nr_osc})
correctly describes the probability error in the presence of
unwanted oscillations of the permanent magnetic field.

\begin{figure}
\includegraphics[width=14cm,height=10cm]{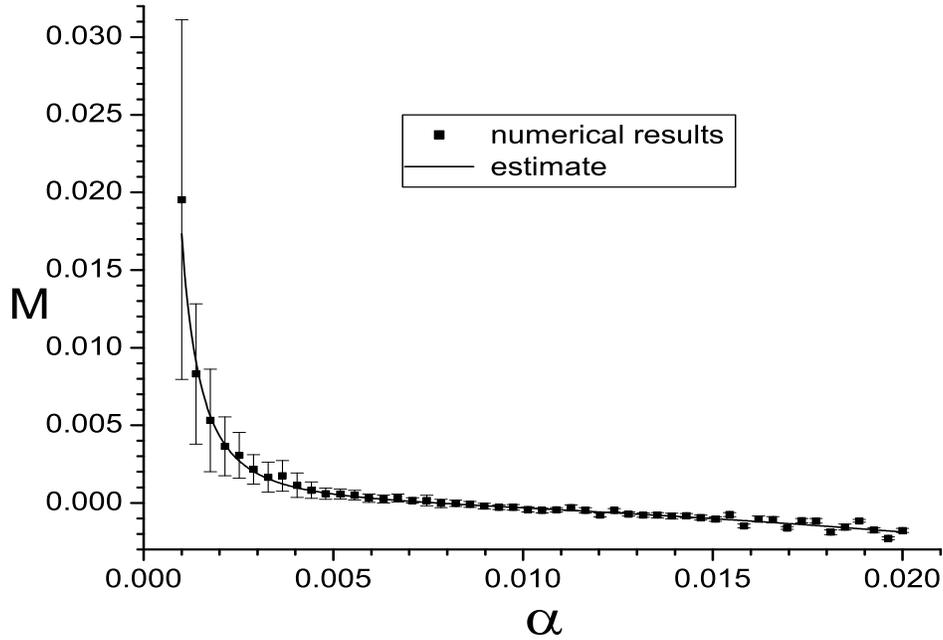}
\vspace{-10mm}
\caption{$M$ as a function of $\alpha$ in the presence of
unwanted oscillations of the external permanent magnetic
field. The parameters are the same as in Fig.~\ref{reffig:10}.
The estimate is calculated using Eq.~(\ref{Mnr_osc}).}
\label{reffig:11}
\end{figure}

Now we will calculate the $z$ component of the magnetization
$M_{\rm osc}$. Consider a typical example with $L=4$ qubits.
If the first pulse of the protocol ($k=1$) generates error
and other pulses do not generate the error,
then after implementation of the entanglement algorithm
there are four states in the register with the following probabilities:
$$
\begin{array}{ll}
|0000\rangle,~~~&{1\over 2}-P_{\rm osc}^\prime(1),\\
|1110\rangle,&P_{\rm osc}^\prime(1),\\
|1111\rangle,&{1\over 2}-P_{\rm osc}^{\prime\prime}(1)\\
|0001\rangle,&P_{\rm osc}^{\prime\prime}(1).
\end{array}
$$
The value of $M_{\rm osc}$ for this superposition is
$$
P_{\rm osc}^{\prime\prime}(1)-P_{\rm osc}^{\prime}(1)+
{L-2\over L}\left[P_{\rm osc}^{\prime\prime}(1)-
P_{\rm osc}^{\prime}(1)\right].
$$
If all the pulses of the protocol generate the error, then the
$z$ component of the magnetization is
\begin{equation}
\label{Mosc}
M_{\rm osc}=P_{\rm osc}(0)+\sum_{k=1}^{L-1}\left[1+{L-2k\over L}\right]
\left[P^{\prime\prime}_{\rm osc}(k)-P^{\prime}_{\rm osc}(k)\right].
\end{equation}
In Fig.~\ref{reffig:11} we compare our estimate
\begin{equation}
\label{Mnr_osc}
M=M_{\rm nr}+M_{\rm osc}
\end{equation}
with the results of numerical simulations for the same values of
$\alpha$ as in Fig.~\ref{reffig:10}. One can see that our estimate
gives us a good approximation of $M$.

\section{Discussion}
We considered the implementation of entanglement in a
two-dimensional ensemble of spin chains.
We demonstrated that the entanglement can be created in the system
with a long-range interaction, such as dipole-dipole interaction,
and the error caused by this interaction can be significantly reduced by
optimization of parameters of the pulses.

If the entanglement is implemented with no error, then $M=0$. We
considered different mechanisms which can generate errors and make
$M\ne 0$. By experimental measurement of $M$ for different
parameters one can define the most important mechanism responsible
for the errors and optimize the design using the obtained
information. The output signal can be enhanced by the multiple copies
of the spin states. The most important dimensionless parameter
characterizing the model is $\alpha$ which is
proportional to the ratio of the Rabi frequency to the difference
$\delta\omega$ between the Larmor frequencies of neighboring
qubits. We now summarize the influence of different kinds of
errors on $M$.

\begin{enumerate}
\item Decoherence. Decreasing $\alpha$ by decreasing
the Rabi frequency one can increase the total time of implementation
of the algorithm and increase $|M|$. The value of $M$ is positive. Decreasing
$\alpha$ by increasing the gradient $|\delta\omega|$ does not
influence $M$.

\item Nonselective excitations (nonresonant transitions).
Decreasing $\alpha$ by decreasing the Rabi frequency or increasing
$|\delta\omega|$ one can decrease $|M|$. The value of  $M$ is negative.

\item Influence of different chains on each other can be estimated
in the following way. (a) One implements the protocol using
the frequency $\nu$ and the Rabi frequency $\Omega$ given by
Eqs.~(\ref{frequency})-(\ref{detuning}). (b) One
takes into consideration the influence of neighboring chains
on each other and modifies the frequencies
and the Rabi frequencies using Eqs.~(\ref{frequency1})
and (\ref{detuning1}). If $|M|$ decreases then the error is mostly caused
by the dipole-dipole interactions between the spins of
different chains.

\item The error is caused by displacements of the qubits if one
observes the following properties: (a) if the gradient
$\delta\omega$ is relatively large [$(v_k/\alpha)^2\gg 1$ in
Eq.~(\ref{Pd})] $M$ is positive and independent of $\alpha$;
(b) the same effect is observed if the distance $a$ between the
qubits is relatively small and the relative displacement
of the $k$th qubit $|v_k|=|dx_k|/a$, where $dx_k$ is the displacement
of the $k$th qubit along the chain, is relatively large, $|v_k|\ge 1/10$;
(c) if the gradient $\delta\omega$ is
relatively small ($|v_k|/\alpha\ge 1$) increasing $\alpha$
we decrease $|M|$ and $M>0$. The latter effect is opposite to
the influence of $\alpha$ in nonresonant excitations where
increasing $\alpha$ we increase $|M|$ and $M<0$.

\item Unwanted fluctuations of permanent magnetic field.
Increasing $\alpha$ decrease $M$ which is positive and $M\sim
1/\alpha^2$.
\end{enumerate}

We did not consider all possible mechanisms such as, for
example, the influence of magnetic impurities in
the substrate~\cite{spinDiffusion}.
Some processes not considered here in detail, such as the
decoherence caused by environment, can be investigated by using
the density matrix if one finds that this kind of decoherence is
the most important mechanism responsible for the errors.

\section*{Acknowledgments}

This work  was supported by the Department of Energy under Contract
No. W-7405-ENG-36 and DOE Office of Basic Energy Sciences.

{}
\end{document}